\documentstyle[twocolumn,seceq,graphicx]{jpsj}

\def\runtitle{Tunneling of Mesoscopic Spins in Molecular Crystals}
\def\runauthor{B. {\sc Barbara}, I. {\sc Chiorescu}, R. {\sc Giraud}, A. G. M.
{\sc Jansen}, A. {\sc Caneschi}}

\title
{
Tunneling of Mesoscopic Spins in Molecular Crystals
}

\author
{ 
B. {\sc Barbara}, I. {\sc Chiorescu}, R. {\sc Giraud},  A. G. M. {\sc
Jansen}$^{1,}$, A. {\sc Caneschi}$^{2,}$}

\inst
{
Laboratoire de Magn\'etisme Louis N\'eel CNRS, BP166, 38042 Grenoble, Cedex-09,
France\\ 
$^1$Laboratoire des Champs Magn\'etiques Intenses, BP166, 38042 Grenoble,
Cedex-09, France\\
$^2$Department of Chemistry, University of Florence, 50144, Italy\\ }

\recdate
{
\today
}

\abst
{The phenomenon of Quantum Tunneling of Mesoscopic Spins is reviewed in the
light of the behavior of the archetype of these systems: the molecular
complex Mn$_{12}$-ac. Most observations can be understood in the framework of
the reduced Hilbert space dimension $2S+1=21$. Due to the large spin
$S=10$, the energy barrier preventing spin rotation is large, and as a
consequence, quantum relaxation is very slow. The application of a magnetic
field of a few Tesla below 1~K allows to observe tunneling (i) between the
states $m=-10$ and $m=10-n$ with $n = 8$ to 11 if the field is longitudinal,
or (ii) between the two ground-states $m\approx-10$ and $m\approx10$ if the
field is transverse. The crossover temperature between ground-state and
thermally assisted tunneling in a longitudinal field extrapolates in zero
field at $\approx1.7$~K. The observation of square root relaxation at
short-times/low-temperatures and of exponential relaxation at
long-times/high-temperatures, as observed previously above 1.5~K, confirms the
important role of the spin bath dynamics which is out of equilibrium in the
first regime and at equilibrium in the second one. In a second part of this
paper a new molecule, so-called V$_{15}$, with resultant spin $S = 1/2$ is
investigated. Contrary to high spin molecules, the energy barrier of low spin
molecules is small or null, and the splitting between the symmetrical and
anti-symmetrical states is sufficiently large to allow spin-phonon
transitions during spin rotation. In low spin molecules the coupling to the
environment is quite different from the one found in large spin molecules in
low fields. }

\kword
{
macroscopic quantum tunneling of magnetic moment, Landau-Zener
mechanism, dissipation, spin-phonon coupling}

\begin{document}
\sloppy
\maketitle

\section{Introduction}

In small ferromagnetic nanoparticles, transitions between the magnetic states
$\pm M$ generally occur by Thermal Activation (TA) over the anisotropy barrier.
These transitions could also occur at constant energy by Quantum Tunneling
(QT), with superposition of the two states $\pm M$. Magnetism being quantum at the
atomic scale and classical at the macroscopic scale, a question raised in the
past was to know what the size of a ``giant spin" should be so that quantum
manifestations could be observed (10, 10$^2$, 10$^3$,$\dots\mu_B$). Early
experiments devoted to this question showed that highly anisotropic rare-earth
intermetallics, like Dy$_3$Al$_2$ and SmCo$_{3.5}$Cu$_{1.5}$, relax almost
independently of temperature below a crossover temperature $T_c$ of a few
Kelvin~\cite{MUBB}. A ``tunneling volume" of about 10$^2$-10$^3$ $\mu_B$ was
obtained (equal to the thermal activation volume at $T_c$). These results and
many others obtained more recently in the 90s on magnetic films and
nanoparticles~\cite{LGBB}, are in agreement with today's expectations for
highly anisotropic systems, but the interpretations were difficult due to the
presence of distributions of energy barriers. The first model describing
thermal and quantum depinning of a narrow domain wall~\cite{Egami}, was
stimulated by the experiments on Dy$_3$Al$_2$~\cite{BB}. Recently, a clear
demonstration of QT was obtained in the large spin molecules Mn$_{12}$-ac, and
later in Fe$_8$. This allows to make a link between magnetism and mesoscopic
physics, which is strengthened by a more recent study on the V$_{15}$  low
spin  molecule. In this paper, the phenomenon of Quantum Tunneling of
Mesoscopic Spins is reviewed in the light of the behavior of the archetype of
these systems: the molecular complex Mn$_{12}$-ac. This system is constituted
of magnetic molecules containing 12 Mn ions with spins $3/2$ and 2,
strongly coupled by super-exchange interactions. The resulting spin $S=10$ is
relatively large and is defined over $\approx1$~nm$^3$, containing 10$^3$
atoms of different species (Mn, O, C, H). The Hilbert space dimension is huge:
10$^8$. However, in the ``mesoscopic quantum tunneling approach" where the
spin $S=10$ is assumed to be ``rigid", a Hilbert space dimension reduced to
21 is sufficient to understand most observations. Such a spin, in both
Mn$_{12}$-ac and Fe$_8$, is large enough to exhibit an important energy
barrier between the states $S=10$ and $S=-10$ in the presence of anisotropy
(of e.g. uniaxial symmetry). This leads to extremely small tunnel splittings
and slow relaxation. The latter can be decreased by orders of
magnitude by the application of a longitudinal or transverse magnetic field. In
Mn$_{12}$-ac, below 1~K, a longitudinal field of a few Tesla allows to observe
tunneling between the states $m=-10$ and $m=10-n$ with $n= 8$ to 11, while a
transverse field of the same order shows ground-state tunneling ($m\approx-10$
to $m\approx10$). The crossover temperature $T_c(H)$ between ground-state and
thermally excited tunneling increases when the longitudinal field decreases.
It extrapolates in zero field to the value $T_c(0)\approx1.7$~K. This value
confirms previous observations of a relatively high crossover temperature in
bulk Mn$_{12}$-ac~\cite{BBjmmm95, BBjmmm98}. Finally, tunneling is deeply
modified by environmental degrees of freedom and in particular by the spin
bath which leads to a square root relaxation at short times.

In a second part of this paper the quantum behavior of a new molecule,
so-called V$_{15}$, containing 15 spins $1/2$, is presented. The resultant
spin $S=1/2$ is small, as a result of antiferromagnetic interactions. Contrary
to high spin molecules, there is no energy barrier and the splitting between
the symmetrical and anti-symmetrical states (given by the matrix element) is
sufficiently large to allow spin-phonon transitions during spin rotation. In
low spin molecules the coupling to the environment may be quite different from
the one found in large spin molecules, unless the latter are with large
non-diagonal matrix elements.

\section{Large spin systems (Mn$_{12}$-ac type)}

The subject of quantum tunneling in molecular crystals started with studies of
the magnetic relaxation of oriented grains of Mn$_{12}$-ac~\cite{Paulsen, Paulsen1,
BBjmmm95, Paulsen2, Novak}. The relaxation time determined above 0.8~K follows an
Arrhenius law $E=kT\ln(\tau/\tau_0)$ with a prefactor
$\tau_0\approx10^{-7}-10^{-8}$~s and an energy barrier close to 60~K in zero
field. Strong deviations to this behavior were observed below 2~K: in a field
tilted by $45^\circ$, the relaxation tends to be independent of
temperature~\cite{Paulsen, BBjmmm95}. This observation was interpreted as QT
of the collective spin $S=10$. Furthermore, relaxation minima and dips in the
ac-susceptibility, observed near zero field, suggested the phenomenon of
resonant tunneling~\cite{BBjmmm95, Novak}. 
\begin{figure}
\includegraphics[width=8.1cm]{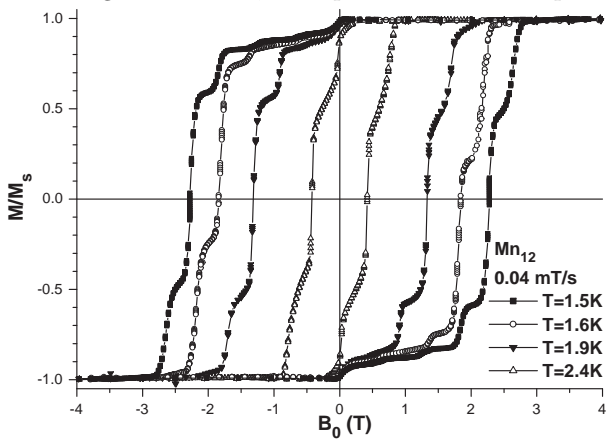}
\caption{Hysteresis loops of Mn$_{12}$-ac, with the field along the $c-$axis.
Alternations of plateaus and steps suggest a ``macroscopic quantization" of the
longitudinal magnetization component. This in fact simply reminiscent of the
quantization of $S_z$ of individual molecules $+$ tunneling in the presence
of a complex environment.} 
\label{fig1} 
\end{figure}
Later, this phenomenon was quantitatively
demonstrated by hysteresis loop measurements on oriented powder~\cite{Fried}
and on a single crystal~\cite{Luc}. Magnetization experiments done in higher
fields and lower temperatures, led to similar conclusions~\cite{Peren}.

\subsection{Tunneling in a longitudinal magnetic field}

For the description of the thermodynamical magnetic properties of 
Mn$_{12}$-ac, the reader could refer to~\cite{jmmm200} and references
therein.
\begin{figure}
\includegraphics[width=8.1cm]{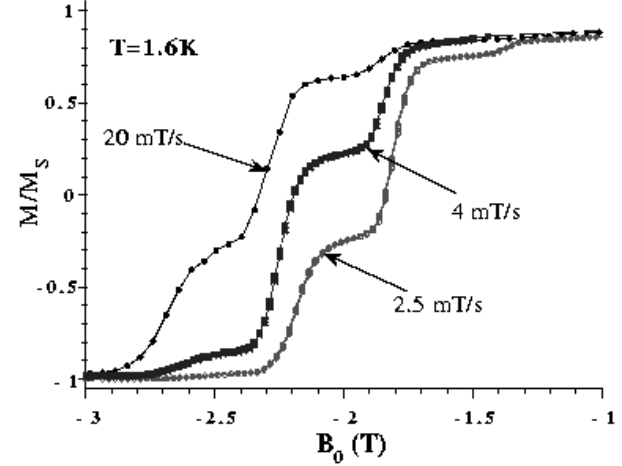}
\caption{Magnetization curves at different sweeping fields in Mn$_{12}$-ac.}
\label{fig2}
\end{figure}

\begin{figure}
\includegraphics[width=6.5cm,angle=90]{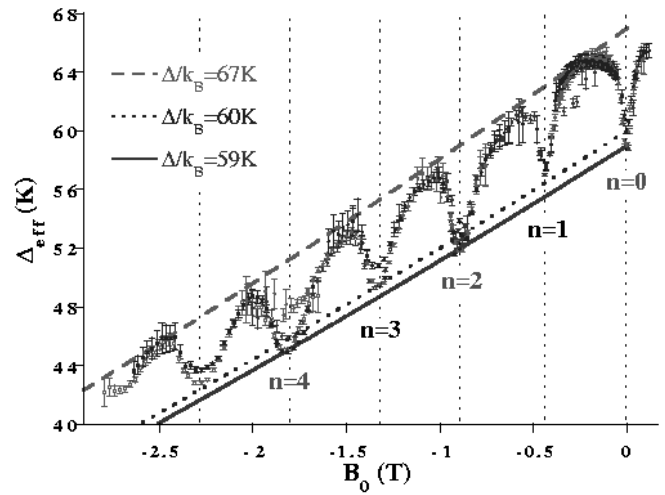}
\caption{Effective energy barrier of Mn$_{12}$-ac vs. a field along the
$c-$axis, showing a parity effect. Values of the zero field barriers are
given.} \label{fig3} 
\end{figure}

The out of equilibrium magnetic properties, are characterized by
isothermal hysteresis loops measured on a single crystal of Mn$_{12}$-ac
(Fig.~\ref{fig1}). The field was applied along the $c-$easy axis of
magnetization with a measuring timescale $\tau_m\approx500$~s per data
point~\cite{Luc}. Below the blocking temperature of 3~K, staircases with
equally separated magnetization steps characterize the dynamics of the system.
In the flat portions of the loops, where the magnetization $M$ has not the
time to relax, its relaxation time $\tau$ is such as $\tau_m\ll \tau$; in the
steep portions of the loop where $M$ relaxes rapidly, $\tau_m\approx \tau$. As
expected, faster sweeping rates give smaller and broader steps (see, in
Fig.~\ref{fig2}, the steps near 1.8~T; note that the last steps are always
broader due to the ``finite size" of the total magnetization). In addition,
relaxation measurements give sharp minima precisely at the fields of the
magnetization steps: $H_n\approx0.44$~n, with $n = 0, 1, 2,\dots$~\cite{Luc}
To illustrate this effect, we give here a dynamical scaling plot
(Fig.~\ref{fig3})~\cite{jmmm200}.
\begin{figure}
\includegraphics[width=8.1cm]{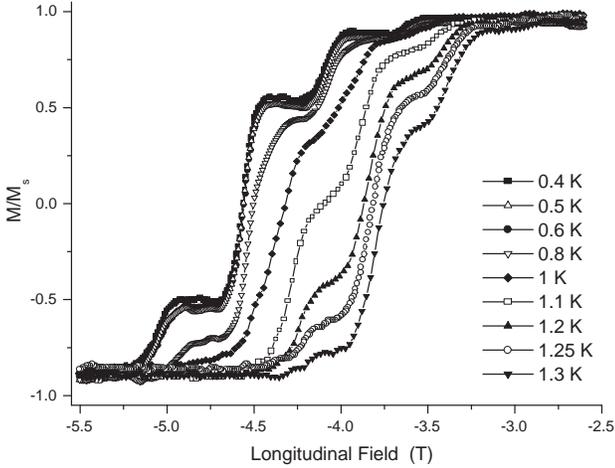}
\caption{Magnetization curves in decreasing field, extracted from torque
measurements with an applied field parallel to the c-easy axis. One observes
that below 0.6~K, the curves are temperature independent, showing that
tunneling is not thermally activated.}
\label{fig4} 
\end{figure}

\begin{figure}
\includegraphics[width=8.1cm]{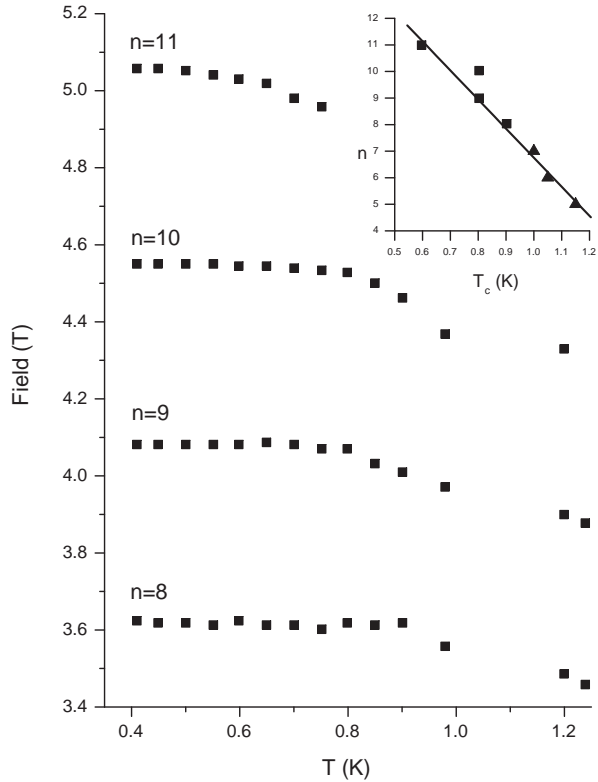}
\caption{Resonance fields $H_n$ measured for $n=8$, 9, 10, 11 at temperatures
between 0.4~K and 1.25~K. For each $n$ a plateau is observed below a critical
temperature $T_c(n)$. This temperature varies linearly with $n$ as shown in the
inset: the triangles are plotted from A. Kent et al.~\cite{Kent} and the
squares were obtained in our study.} 
\label{fig5} 
\end{figure}

Fig.~\ref{fig4} shows several hysteresis loops, recently measured at the High
Magnetic Field Laboratory in Grenoble (LCMI), using a torque magnetometer below
1.3~K~\cite{IC}. The way to extract the longitudinal magnetization (parallel
to the easy $c-$axis) from the torque, will be published elsewhere and
compared to previous experiments of Perenboom et al~\cite{Peren}. In
Fig.~\ref{fig4}, the transverse component of the applied field was smaller
than 0.1~T. The obtained hysteresis loops show magnetization jumps at
certain values $H_n$ of the longitudinal field, as in Fig.~\ref{fig1}, but the
temperature being here rather low, the entire hysteresis loops become
independent of temperature below a crossover temperature $T_c\approx0.6$~K.
This low temperature saturation, already observed\cite{Peren}, suggests that
thermal activation is not relevant at these temperatures, and that tunneling
takes place from the ground-state $m=-10$ to a state $m=10-n$, with $n = 8$,
9, 10 and 11, depending on the exact value of $H_n$. These
fields become independent of temperature below
$T_c(n)$ but they decrease above this temperature (Fig.~\ref{fig5}),
confirming the existence of a crossover at $T_c$ (in the presence of fourth
order crystal field terms). The inset of Fig.~\ref{fig5} shows that $T_c$
increases almost linearly when $n$, i.e. the field, decreases, as expected.
Interestingly, a linear extrapolation of $T_c$ vs. $n$ gives the zero-field
crossover temperature $T_c(0)\approx1.7$~K. This value is consistent with the
temperature of 1.9~K, below which the beginning of a relaxation plateau was
observed~\cite{BBjmmm95, BBjmmm98, LucPRL}, confirming the negligible role of
the minority phase of Mn$_{12}$-ac at long time-scales~\cite{jmmm200, LucPRL}. A
more detailed description of these experiments and their interpretation will
be given in a further report.

\subsection{Tunneling in a transverse magnetic field}

\begin{figure}
\includegraphics[width=8.1cm]{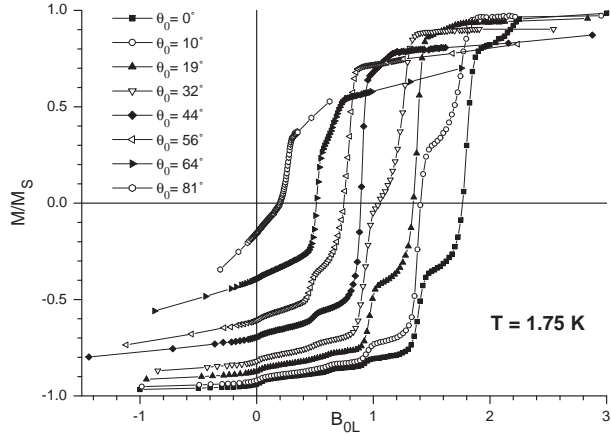}
\caption{Magnetization curves measured in Mn$_{12}$-ac with a tilted magnetic
field (the angles are with the $c-$axis).} 
\label{fig6}
\end{figure}

In order to study the effect of a field perpendicular to the easy axis of
magnetization, hysteresis loop measurements were first performed above 1.5~K
with a field tilted by an angle $\theta$~\cite{FredJAP, jmmm200}. As for
$\theta=0$, alternations of plateaus and steep variations of the magnetization
$\Delta M$, are observed (Fig.~\ref{fig6}). $\Delta M$ increases with the
transverse field component $H_T = H\sin\theta$, but the positions of the steps
remain at the same longitudinal field $H_{Ln} =
H\cos\theta\approx\dfrac{nD}{g\mu_B}$. Longitudinal and transverse fields
having, to first order, ``orthogonal" effects (the former puts the levels in
coincidence and the latter removes the degeneracy of these coincidences by
creating the tunnel splitting), it is normal that a transverse field does not
modify significantly the coincidence if the applied field is much smaller than
$H_A$, but only increases the tunneling probability. The measured
relaxation times follow an Arrhenius law with a barrier reduced by the
transverse field. The observed increase of $\Delta M$ with $H_T$ and $T$ is 
due to (i) easier TA resulting from the lowering of the energy barrier as
predicted by the classical expression $\propto1-2(H_T + H_L)/H_A$ for
$(H_T + H_L)\ll H_A$ and/or (ii) easier QT resulting from the increase of the
tunnel splitting of resonant level pairs $-m$ and $m-n$,
$\Delta\propto(H_T/H_A)^{2m-n}$ + higher terms. A comparison between
these two contributions shows that thermally activated tunneling on excited
states takes place on deeper and deeper energy levels when $H_T$ increases.
The transmission rates of different thermally assisted channels in a transverse
field were calculated ~\cite{jmmm200} and compared to experiments. 

\begin{figure}
\includegraphics[width=8.1cm]{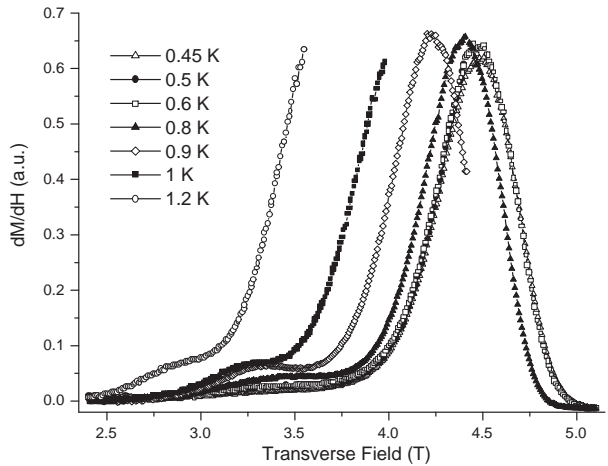}
\caption{Variation of $dM/dH$ vs. $H$, measured in Mn$_{12}$-ac with a
transverse field of about 4~T. Below $\approx0.6-0.7$~K, the curves are
temperature independent.} 
\label{fig7} 
\end{figure}

It was concluded that ground-state tunneling is obtained for $H_T$ larger
than a few Tesla. This effect was recently clearly shown for the first time.
For that we made the same sub-Kelvin torque experiments as above, but in the
presence of a field almost perpendicular to the easy $c-$axis of
magnetization~\cite{IC}. In this case the symmetry of the double potential
well of Mn$_{12}$-ac is essentially preserved and ground-state tunneling
between the states $m\approx-10$ and $m\approx10$ can be observed. In these
experiments, the tilt angle was very close to $90^{\circ}$, within a range of
one degree, giving a transverse field component between 3.5 and 4.5~T, and a
longitudinal component varying around 0.04~T. Fig.~\ref{fig6} shows the $n=0$
resonance, observed in zero internal field. The longitudinal component of the
applied field is exactly opposite to the Lorentz field, to cancel it. In order
to show this resonance more clearly at different temperatures below 1.3~K, we
plotted in  Fig.~\ref{fig7} their profile $dM/dH$ vs. $H$. As observed in
longitudinal fields, the curves are independent of temperature below a certain
temperature, which is here $\approx0.6-0.7$~K. This shows that below
$T_c\approx0.6-0.7$~K the tunneling, observed in zero internal field, is of
pure quantum origin and corresponds to $m\approx-10$ to $m\approx10$
transition. In particular phonon emissions which must be present in the case
of asymmetrical potential wells are absent here and the relaxation plateau
cannot be attributed to  self-heating. Besides, thermalization is optimal in
the $^3$He bath of these experiments. Note that the $m=-10$ and $m=10$
ground-states are in fact slightly mixed by the transverse field, and the
$z$-component of $\mib S$ in each well is reduced to a value which can be
approximated by $m = S_z = \pm S\cos\phi\approx9$ which is nevertheless close
to 10 ($\phi$ is the angle between $\mib S$ and the easy $c-$axis).

This first clear observation of ground-state tunneling of the main phase of
Mn$_{12}$-ac allows to study the quantum relaxation of this system (next
section). Note that in all these experiments with longitudinal and transverse
fields of several Tesla, the minority phase of Mn$_{12}$-ac is
superparamagnetic and cannot contribute to hysteresis loops and relaxation
experiments, unless by increasing on a negligible way the dynamics through
weak dipolar interactions.

\section{Resonant tunneling mechanisms}

\subsection{General interpretation}

The simplest way to interpret these results is to calculate the spectrum of
energy levels in the presence of a magnetic field. For that we use the
following Hamiltonian for a spin $\mib S$ in a crystal field of
tetragonal symmetry, with an applied field $\mib H(H_x,H_z)$. This Hamiltonian
contains a diagonal part $H_1$ and a non-diagonal $H_2$ one :
\begin{equation}
H_1 = -DS_z^2 -BS_z^4-g\mu_BS_zH_z  
\label{H1}
\end{equation}
and 
\begin{equation}
H_2 = -C(S_+^4+S_-^4)-g\mu_BS_xH_x .
\label{H2}
\end{equation}
In so doing, one neglects all the internal degrees of freedom
of the collective spin, and this is precisely the interest of this
``mesoscopic" approach. With 4 spins $3/2$ and 8 spins 2, the dimension of the
Hilbert space of the collective spin of Mn$_{12}$-ac is 10$^8$ (or 6$^8$ for
Fe$_8$ and 2$^{15}$ in V$_{15}$). Together with extremely small gaps, this very
large dimensions can be taken as a characteristic of mesoscopic spins.
Nevertheless we expect some effects associated with the fine structure of the
molecule level in connection with e.g. intra-molecular anti-symmetric
Dzyaloshinsky-Moryia interactions $\mib D_{ij} \cdot(\mib S_i\times \mib S_j)$,
which do not conserve $\mib S^2$ ~\cite{jmmm200, BBjmmm98}. We assume that
these terms are more or less well represented by different contributions in
Eq.~\ref{H1},~\ref{H2}. The energy spectrum in longitudinal field is given by
the hamiltonian $H_1$, $E(m) = -Dm^2-Bm^4-g\mu_BmH_z$ where $m=S_z$. The
levels $m (>0)$ and $n-m (<0)$ cross when $E(m) = E(n-m)$. The crossing
fields (or resonance fields) are given by:
\begin{equation}
 H_n = \dfrac{nD}{g\mu_B}[1+\dfrac{B}{D}((m-n)^2+m^2)]  .
\end{equation}
If the effects of $B$ were neglected, the crossing fields would be equally
separated and given by $H_n\approx nD/g\mu_B = nH_A/(2S-1)$, where $n =1,
2,\dots$ The comparison with the observed sequence of fields $H_n\approx0.44n$
gives $D/g\mu_B\approx0.44$~T i.e. $D\approx0.60$~K. The value
$B/D\approx0.002$ shows that the shift due to $B$ is small for excited levels
but rather large for lowest levels. In the following we will take the set of
values $D=0.56$~K, $B=1.1$~mK, $C= 3\cdot10^{-5}$~K obtained in EPR~\cite{BarraPRB}.
In low fields above 2~K, $\Delta H_n\approx0.41(1 + 0.004m^2)$ suggests that
tunneling occurs mainly through $m\approx4$. Energy barrier determination
from relaxation measurements on the magnetization plateau near zero field gives
$\vert E \vert \approx67$~K. This is also in agreement with these parameters:
for $n=0$, $E=-D_{eff} S^2$ with $D_{eff}= D + B S^2$  gives $E\approx56 +
10^{-3}S^4\approx67$~K (for $S=10$). Finally, direct anisotropy field
measurements (magnetization measurement in a field perpendicular to the easy
axis) give $\vert E \vert \approx65$~K~\cite{FredJAP,BBjmmm98} which is close
enough to the EPR value of 67~K~\cite{BarraPRB}.

The non-diagonal part of the hamiltonian $H_2$ removes the degeneracy at
level crossings giving the splitting $\Delta$ (anti-crossing or
level-repulsion). Following van Hemmen and S\"ut\H o~\cite{Hemmen}, we can
write for a non-diagonal term $C_p$ of order $p$:
\begin{equation}
\Delta(S,0)= \frac{DS}{\pi}\left(\frac{2DS^{2-p}}{C_p}\right)^{-2S/p} 
\label{deltaS0}
\end{equation}
The gap is an exponentially small fraction of the anisotropy level separation
$\approx DS$ and is in general extremely small for large spins. This
expression can be extended to $n\not=0$:
$\Delta_{(m,n)}=(Dm/\pi)[(2D/C_p)\cdot S^{(2-p)}\cdot f(m/S)]^{-(2m-n)/p}$
where $f(m/S)$ is a function which could be evaluated. The gaps $\Delta$
associated with the resonances allowed by  the anisotropy $(-m, m'=m-n, n)$
were calculated numerically in Mn$_{12}$-ac, by exact diagonalization of the
$S=10$ matrix for the parameters given above (see Table~\ref{table:1}). Only
those transitions such as $m-m\prime$ is a multiple of 4 are allowed. If a
transverse field is also taken into account in the diagonalization, the
condition becomes simply $m-m\prime =$ integer.
\begin{table}
\caption{Non-zero tunnel splittings calculated by exact diagonalization for the
spin $S=10$ of Mn$_{12}$-ac. Each cell contains the splitting in Kelvin and the
label of the crossing levels under the form $(-m, m', n)$. The crystal field
parameters~\cite{BarraPRB} are given in section 3.1.}
\label{table:1} \begin{tabular} {| @{$\;$}c| @{$\;$}c| @{$\;$}c| @{$\;$}c|
@{$\;$}c|} \hline
(-10,10,0)& (-10,6,4)&(-10,2,8)& (-10,-2,12)& (-10,-6,16)\\
$1.15\cdot10^{-11}$& $4.67\cdot10^{-8}$&$1.61\cdot10^{-5}$&
 $1.84\cdot10^{-3}$& $9.67\cdot10^{-2}$ \\ \hline  
 (-9,7,2)& (-8,8,0)& (-8,4,4)& (-8,0,8)& (-8,-4,12)\\ 
 $9.41\cdot10^{-8}$& $1.16\cdot10^{-7}$& $7.45\cdot10^{-5}$&
 $8.68\cdot10^{-3}$& $2.98\cdot10^{-1}$  \\ \hline 
 (-9,3,6)& (-7,5,2)& (-6,6,0)& (-6,2,4)& (-6,-2,8)\\ 
 $4.19\cdot10^{-5}$& $1.03\cdot10^{-4}$& $1.14\cdot10^{-4}$&
 $1.77\cdot10^{-2}$& $4.97\cdot10^{-1}$ \\ \hline 
 (-9,-1,10)& (-7,1,6)& (-5,3,2)& (-4,4,0)& (-4,0,4)\\
 $4.69\cdot10^{-3}$& $1.33\cdot10^{-2}$& $2.08\cdot10^{-2}$&
$2.2\cdot10^{-2}$& $6.38\cdot10^{-1}$ \\ \hline 
 (-9,-5,14)& (-7,-3,10)& (-5,-1,6)& (-3,1,2)& (-2,2,0)\\
 $1.94\cdot10^{-1}$& $4.02\cdot10^{-1}$& $4.97\cdot10^{-1}$&
 $6.76\cdot10^{-1}$& $6.89\cdot10^{-1}$\\ \hline
\end{tabular}
\end{table}
 As an example, $H_x=0.1$~T gives $\Delta/k_B=2$~mK for $m=m\prime=3$.
The fast increase of $\Delta$ with $n$ explains why a longitudinal field
allows to observe the relaxation of Mn$_{12}$-ac below 1~K. The gaps also
increase dramatically with a transverse field as shown by Eq.~\ref{deltaS0}.

\subsection{Tunneling probability between two levels}

If the two levels were exactly in coincidence the tunneling rate 
would simply be given by $\Delta/h$. However this situation never occurs
because levels are never in ``static coincidence". If one assumes that a
magnetic field is applied with the rate $c=dH/dt$ at the anti-crossing of two
levels, the time $\Delta/cg\mu_BS$ during which the field crosses the region
$\Delta/g\mu_BS$ where the magnetic states are mixed, can be much smaller or
much larger than the characteristic oscillation time of the two-level system 
$h/\Delta$. In the first limit, the initial spin state $S_z = -m$ has not
enough time to mix with $S_z = m-n$ and it is unchanged (non adiabatic
transition). In the second one, $-m$ changes in $m-n$ and at the anti-crossing
the two states are mixed giving oscillations of $S_z$ (adiabatic case). In
this intermediate region, $S_z =0$. In the general case~\cite{Miyashita, LZ,
LGunther, Dobro} (Landau-Zener model) the transition probability is given by
the ratio of these two times:    
\begin{equation} 
P_{m,n} = 1-\exp\left({-\Gamma_{m,n}}\right)   
\end{equation}
with  
\begin{equation}
 \Gamma_{m,n}=\pi\Delta_{(m,n)}^2/[2\hbar (2m-n)g\mu_BdH/dt]
\end{equation}
If $\Gamma_{m,n} \ll1$, $P_{m,n}\approx\Gamma_{m,n}$. In large spin systems
such as Mn$_{12}$-ac or Fe$_8$ where zero-field ground-state
splittings $\Delta/g\mu_BS$ are of the order of 10$^{-9}$ and 10$^{-11}$~T, the
regions $\Delta$ are  crossed in 10$^{-9}/c$ to 10$^{-11}/c$  seconds. The
oscillation time $h/\Delta$ is going from  10$^{-3}$ to 0,1 second, the
Landau-Zener probabilities are $\Gamma\approx10^{-6}/c$ and 10$^{-10}/c$,
(where $c$ is expressed in T/s). Typical values of $c$ (between 10$^{-3}$ and
10$^3$~T/s) give $\Gamma\ll1$ i.e. essentially non-adiabatic Landau-Zener
transitions (in the absence of fluctuations and magnetic field). It is clear
that with an applied field, $\Delta$ can reach values approaching $P\approx1$
(Eq.~\ref{deltaS0}), and this is the reason why ground-state tunneling can
easily be observed in Mn$_{12}$-ac, as shown above.  

\subsection{Effects of temperature}

For a given coincidence field ($n$ fixed), $-m$ can take different values $-m =
-10, -9,\dots$ up to the top of the barrier. A tunneling channel can be
associated with each one of  these values of $-m$, giving paths such as
$-S\Rightarrow -m\Rightarrow m \Rightarrow S$. The first portion
($-S\Rightarrow -m$) is thermally activated while the second one
($-m\Rightarrow m$) is tunneling. The third one ($m \Rightarrow S$) is
associated with phonon emissions. The total transition rate is given by the
product of the Thermally Activated rate $\Gamma (m) =  \tau_0^{-1}\exp{(E(-S) -
E(-m))/kT}$ by the tunneling probability $P_{m,n}$ between $-m$ and $m-n$:   
\begin{equation}
\Gamma_{TA} = P_{m,n} \tau_0^{-1}\exp{(E(-S) - E(-m))/kT} 	 	
\end{equation}
This expression explains the Arrhenius behavior of the relaxation time in the
high temperature regime of Mn$_{12}$-ac. In this regime, tunneling simply
modifies the prefactor of the Arrhenius law (whose temperature dependence is
weak compared to the exponential). Depending on the values of temperature and
field, three different tunneling regimes can be defined:

 $-$ The \emph{thermally activated tunneling regime}, where tunneling takes
place on upper paths. These paths are on levels $m<m_c$, for which the
tunneling probability is much larger than the thermal one, i.e.
$P_{m,n}\approx1$. This high temperature regime allows to understand
the passage to the classical behavior. The rate for $-m$ to $m-n$,  change is
given by:
\begin{equation}
\Gamma_{Ac}\approx \tau_0^{-1}\exp{(E(-S) - E(-m_c))/kT}  
\label{GammaAc}  
\end{equation}
Above $E(m_c)$, the barrier is transparent: tunneling is so fast that it
short-circuits the top of the barrier at energies larger than $E(m_c)$. One
can see in Eq.~\ref{deltaS0} that for small $m$, $\Delta$ becomes comparable to
crystal field level separations and everything is so well mixed that the
states of spins up and down have no meaning (there is no barrier). In the
presence of a longitudinal field, one can also define an ``effective barrier":
\begin{equation}
E(m,n)=E_{eff}(H)\approx E_{eff}(0)(1-H/H_A)^\alpha, 
\label{barrier}
\end{equation}
with 
\begin{eqnarray}
E_{eff}(0) = E(-m_c)-E(-S) \approx D_{eff}(S^2-m_c^2) \nonumber
\end{eqnarray}
The notion of effective barrier $E_{eff}(H)$ has the advantage to provide a
simple scaling expression (Eq.~\ref{GammaAc} and Eq.~\ref{barrier}) for the
analysis of relaxation experiments in the thermally activated tunneling
regime. The relaxation times $\tau(H,T)$ measured in Mn$_{12}$-ac between 2.5
and 3~K have been analyzed using the scaling plot of $E_{eff}=T\ln \tau(H,T)$,
with $\alpha\approx2$ or $3/2$ (Fig.~\ref{fig3}). The effect of tunneling is
equivalent to a reduction of energy barrier from 67~K to 60~K (in zero field),
i.e. of about $10\%$. A comparison with (Eq.~\ref{barrier}) gives
$(m_c/S)^2\approx0.1$, and this shows that tunneling short-circuits the top of
the barrier when $m \le m_c\approx3$. Interestingly, ac-susceptibility
experiments give $E\approx64$~K~\cite{jmmm200, FredJAP}, and this is a larger
value than the 60~K of quasi-static experiments because ac-experiments are
faster, giving smaller reduction of the top of the barrier.  This scaling
analysis is more accurate than previous plots. It shows dips in
$E_{eff}$ which are more important for even values of $n$ than for odd ones.
This parity effect seems to be a tunneling manifestation of the fourth order
transverse anisotropy terms $C(S_+^4 + S_-^4)$ which contributes to tunneling
only for $n$ even or multiple of 4 (depending on the parity of m). In fact it
is very general, in particular because it occurs not only for fourth order
terms but also for other orders, such as the order one if the timescale is
short enough~\cite{Miyashita, Garanin}. As an example, in the presence of a
transverse magnetic field, even resonances are induced by $S_+^4$ or $S_-^4$,
whereas odd resonances are induced by combinations like by $S_xS_+^4$ or
$S_xS_-^4$~\cite{Igor}. It is clear that these effects, observed here at high
temperature, also exist at low temperature.

$-$ The \emph{thermally assisted tunneling regime} is for intermediate paths. 
The TA probability $\propto \exp({-D_{eff}(S^2-m_c^2)/kT})$ and the tunneling
one  are not very different from each other. This is for $m > m_c\approx3$,
but not too large, otherwise levels are too deep and tunneling is too slow. 
The thermally assisted tunneling rate for $-m$ to $m-n$ transitions can be
written: 
\begin{equation}
\Gamma_{As}\approx \Gamma_0 \exp{(D_{eff}(S^2-m_c^2))/kT},
\label{GammaAs}
\end{equation}
with
\begin{eqnarray}
\Gamma_0 = \dfrac{\pi \Delta^2}{2\hbar g\mu_B(2m-n)h_s}\frac{\tau_s}{\tau_0} \nonumber 
\end{eqnarray}
if $\tau_0 \ll \tau_s$, where $h_s$ and $\tau_s$ are the amplitude and the
characteristic time of internal field fluctuations, and $\tau_0$ the prefactor
of the Arrhenius law. In the presence of dipolar field distribution $\Gamma_0$
is multiplied by $h_s/H_D$ where $H_D$ is the width of the
distribution~\cite{ProkStamp}. The tunneling gap can be evaluated by a
comparison of  measured and calculated relaxation rates (Eq.~\ref{GammaAs}).
In assisted tunneling regime, where $\tau\approx10^6$~s, at $T\approx2$~K,
$H\approx0$, $\tau_0\approx10^{-8}$~s and $H_D\approx40$~mT  and
$\tau_s=T_1\approx10^{-4}$~s (see below), we find $\Delta/k_B\approx3$~mK. This
value is close to the gap $\Delta/k_B\approx2$~mK, calculated for $m =
m_c\approx3$ (Table~\ref{table:1}). For $m > 3$ or 4, $\Delta$ decreases
dramatically (Eq.~\ref{deltaS0} or Table~\ref{table:1}), and relaxation times
become very long ($\Gamma_0 \propto \Delta^2$). In order to observe
ground-state tunneling it is necessary to increase level mixing by applying a
magnetic field. 

$-$ The \emph{Ground-State tunneling regime}. In the presence of a
longitudinal field, the tunneling path $-S$ to $+S-n$ is followed by phonon
emissions while $+S-n$ reaches the ground-state $+S$. In the absence of
longitudinal field ($n=0$), tunneling occurs between $-S$ and $+S$ 
(ground-state tunneling). Energy conservation is naturally fulfilled, but
couplings to the environment are still required for angular momentum
conservation. In both cases tunneling rate can be enhanced by the application
of a transverse field. At $T=0$ the tunneling rate is given by (for $\Gamma
\ll1$):   \begin{equation}
\Gamma_{GS}=P_{(-S,S-n)}/\tau_s=\pi\Delta_{(S,n)}^2/[2\hbar g\mu_B(2S-n)H_D]
\end{equation}	   
where is $H_D$ the internal dipolar field ($g\mu_BSH_D =E_D$). Here too the
comparison between measured and calculated relaxation times gives access to
the tunnel splitting. The relaxation time measured in zero field at 1.8~K
(beginning of the plateau), $\tau\approx10^9$~s~\cite{BBjmmm95, BBjmmm98,
Fried}, gives $\Delta/k_B\approx10^{-10}$~K~\cite{jmmm200}. This value is not very
much different from $\Delta/k_B\approx10^{-11}$~K, calculated by diagonalization
for $S=10$ (Table~\ref{table:1}). A value of $C\approx4.5\cdot10^{-5}$~K
instead of $3\cdot10^{-5}$~K could explain this difference. 

Using more recent relaxation experiments below 1.3~K and in the presence of a
transverse field of 4~T, we find a tunnel splitting of
$\Delta/k_B\approx 10^{-5}$~K (section 4.3).

\subsection{Effects of environment}

The crucial role of couplings to the environment in ``Macroscopic Quantum
Tunneling" was pointed out several years ago~\cite{Leggett}. These
couplings tend to destroy MQT effects, but at the same time they are necessary
to observe it. Environmental couplings giving dissipation, the spin motion
between the two wells is non-equally damped, and this dephases the two wave
functions and destroys their interferences: the system becomes more classical
and tunneling is eventually suppressed. In Ref ~\cite{jmmm200} we gave a
physical picture of this effect, called ``topological
decoherence"~\cite{Stamp1, StampPRL88}, in terms of the Berry
phases~\cite{Berry}.

Tunneling can also be suppressed if the environment contains internal bias
fields, which brings the molecules out of coincidence. The main origins of
internal fields are the quasi-static components of hyperfine fields of the
molecule itself and the quasi-static components of dipolar fields between
molecules. They give a Zeeman energy distribution for molecular spins. The
Zeeman shifts of the states $+m$ and $-m$ of a molecule with non-zero internal
field, can be compensated by the application of an external field of opposite
value. However this compensation cannot be exact due to finite field
resolution $\delta H$, (unless $g\mu_BS\delta H < \Delta$, which is unrealistic
for mesoscopic spins). This basic interdiction for tunneling can be lifted
by the environment, and in particular by fast fluctuations of internal fields.
If the energy separation of the levels $+m$ and $-m$, is smaller than
$g\mu_BmH_f$ where $H_f$ is the amplitude of the fluctuating field, these
levels will come into coincidence $1/\tau_f$ times per second, where $\tau_f$
is the mean fluctuation time, allowing tunneling at each coincidence in the
``energy window " $g\mu_BmH_f$. This effect of dynamical (homogeneous)
broadening ``averaging" static inhomogeneous broadening, in a given energy
window was recognized only recently~\cite{ProkStamp}. In this work the
fluctuations were assumed to be of hyperfine origin:  the fast fluctuations of
amplitude $H_f$ are those of the hyperfine fields and their characteristic
time is $\tau_f = T_2$. However at large enough temperature, the temperature
dependent process $T_1$ or fast  fluctuations of the dipolar fields can also
be relevant.

Note that the applied field itself is never completely stable and its
fluctuations about a stable mean value can induce tunneling. This seems to be
an interesting aspect of the measurement in quantum mechanics, where
fluctuations in the measuring apparatus induce tunneling (here the coils, the
current supply). In an ideal system where the environment would only be
constituted of the measurement set up without any other fluctuations 
tunneling could only be observed through the imperfections of the measuring
tool. The width of the tunnel transition would be simply given by the
instrumental homogeneous broadening of quantum levels, allowing tunneling in
the window $g\mu_Bm\sqrt{<h^2>}$, where $\mib h (t)$ is the fluctuating part
of the field $\mib H$ conjugated to the order parameter. Further analysis will
require to distinguish between our ability to select a given value of the
applied field and the uncertainty on the determination of this value. Note
that one may simulate a ``bad measuring tool", by applying a fluctuating field
of amplitude larger than the internal fields. The simplest case, which we use
commonly, is to apply an ac field to accelerate the dynamics, but the
simulation of a ``bad apparatus" would require to apply a field random in time
(e.g. white noise) and to test the effect of different statistics on the
resultant tunneling effect (amplitude, time-scale, white noise or with
correlations…).

In systems such as Mn$_{12}$-ac or Fe$_8$, hyperfine and dipolar field
fluctuations are in general more important than those of the measuring tool.
When the temperature is such that the first excited level $m=9$ is
occupied (roughly above 1~K), fast fluctuations come from local dipolar and
hyperfine fields $H_{loc}$ around their mean values, with $\Delta I=\Delta
m=\pm1$ co-flips ($T_1$), and also at higher temperature from phonon
transitions associated with $\Delta m=\pm1$. At low temperature when these
process are frozen, only $T_2$ remains. The dominant fluctuation time given by
an Orbach process~\cite{Villain} is $T_1\approx\tau_0\exp(E(m)-E(S))/k_BT$
and $H_f\approx  H_{loc}/S\approx H_{loc}/I\sqrt N$. For Mn$_{12}$-ac at about
2~K, we get $T_1\approx10^{-4}-10^{-5}$~s and $H_f\approx1-5$~mT ($S=10$,
$I=5/2$ and $N=12$ for the Mn nuclear spins in a molecule). NMR on Mn nuclear
spin was recently made possible by Goto et al~\cite{Goto}. Transfered
hyperfine fields of protons give $T_1\approx10^{-4}$~s and
$H_f\approx1.4$~mT~\cite{Borsa}.  

\section{Relaxation laws and resonance profile}

There are two ways to describe a resonance. In the first one, one plots the
profile of the resonance determined at a given time-scale, e.g. by plotting
$dM/dH$ vs. $H$, for a given sweeping field rate $dH/dt$. In the second one,
the field is kept constant and one plots the magnetization vs. time, $M(t)$.
Both ways were studied in Mn$_{12}$-ac. It was shown that relaxation laws
$M(t)$ and resonance profile $dM(H)/dH$ are not independent.   	

\subsection{Thermally activated relaxation}
In the high temperature regime, above $\approx2.5$~K i.e. not far from $T_B$,
the tunnel splitting $\Delta$, averaged over a few excited levels, is of the
order of the resonance line-width (about 0.5~K). In this case the
resonance, homogeneously broadened by spin-phonon transitions of energy 
$\Delta$, has a Lorentzian line-shape and the relaxation is exponential
($M\propto\exp(-t/\tau)$). The contribution of dipolar fields to the resonance
line-width is also Lorentzian, because any ``reaction" of dipolar fields $H_D$
to spin reversals, is continuously averaged in time, meaning that the system
is equilibrated. The tunnel window covers the whole resonance and since
$g\mu_BH_D\approx\Delta$, the resonance width is of both origins (dipolar and
intrinsic i.e. with spin-phonon transitions). This equilibrated regime,
corresponds to what we called above the ``thermally activated tunneling"
regime. 

\subsection{Thermally assisted relaxation}
In the intermediate temperature regime (below $\approx2.5$~K), dipolar fields
are essentially frozen leading to an  inhomogeneous distribution of internal
fields. The intrinsic tunnel window of width $\Delta$ and the
small fluctuating parts of internal fields are much narrower than the width of
the inhomogeneous distribution of internal fields: the latter is not averaged
by the quantum dynamics and its profile could be anything (e.g. Gaussian in
particular theoretical situations, but numerical simulations showed that it is
in general more complex than either Lorentzian or Gaussian). In Mn$_{12}$-ac
at 1.6~K, resonance peaks are shifted to lower fields and their width
increases after zero-field cooling, showing that the inhomogeneous
distribution of dipolar fields is at this temperature broad enough to
determine the resonance width~\cite{jmmm200}).

In this intermediate temperature regime, strong deviations from exponential
relaxation were observed, and in particular a square root law~\cite{LucLowT,
LucPRL}.
\begin{figure}
\includegraphics[width=8.1cm]{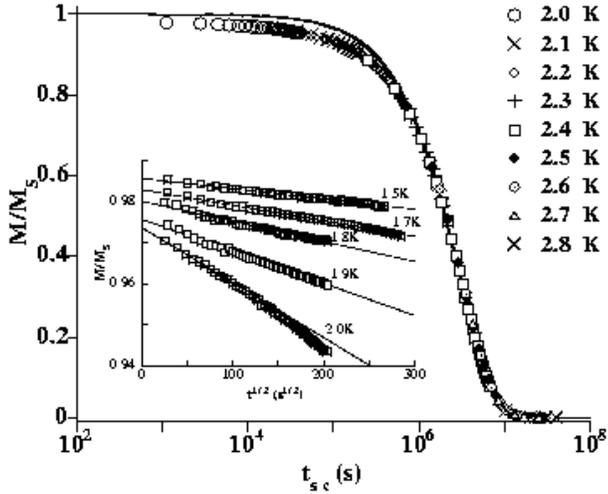}
\caption{Scaling plot of the relaxation in Mn$_{12}$-ac in zero applied field.
Strong deviations from an exponential occur below $t_{sc} \approx10^6$~s
(continuous curve). The scaling variable is $\tau_{sc} =
(t+t_w)\cdot \tau(2$~K$)/\tau(T)$, where $\tau$ is the relaxation time and
$t_w$ is an experimental waiting time. Inset: low temperature regime where the
relaxation is nearly a square root of the time.}  
\label{fig8}  
\end{figure}

However this law is not limited to 0~K and $M\approx M_s$ as
predicted. Furthermore a crossover between square root and exponential
relaxation is quite apparent in Mn$_{12}$-ac. A scaling analysis
(Fig.~\ref{fig8}) shows (i) an asymptotic exponential regime between 3~K and
$\approx2.7$~K (high-temperatures / long-times), (ii) important deviations
from this regime between $\approx2.5$ and 2~K, and finally (iii) a square root
asymptotic regime below 2~K (low-temperatures / short-times). This low
temperature regime was recently extended down to 0.4~K (see below).  The
crossover region (ii) is broad and not trivial. In particular it shows highly
non-linear S-shaped $M(t)$ curves. Furthermore, it takes place at temperatures
where Mn$_{12}$-ac should be equilibrated, with exponential relaxation. These
observations seem to result from the decrease of the total magnetization
$M(T,t)$, with increasing time or/and $T$~\cite{jmmm200}. In this case, the
mean internal field decreases, and this leads to opposite shifts of the
spin-up and spin-down density of states. The first resonance (in zero internal
field) is maximum when the applied field compensates the mean internal field
(Lorentz field): $H_0 = -H_{int} = -CM(T,t)$, where $C= 4\pi/3 - N$ gives the
difference between the Lorentz field and the sample demagnetizing field. If
the sample has a needle shape, as in Mn$_{12}$-ac, this field $H_1\approx-4\pi
M(T,t)/3$, is negative. In other systems with different shapes, like Fe$_8$,
it is positive. In the course of relaxation the $+S$ density of states
decreases (due to the decrease of $H_1(t,T)$) and this gives also a square
root relaxation, but which is valid at high temperature i.e. in the thermally
assisted regime (at low temperatures where the changes in $M(t,T)$ are too
small, the square root regime is exclusively described in the frame of
model~\cite{ProkStamp} ). Furthermore the relaxation law may depend on whether
the relaxation is performed exactly at resonance or not. This allows in
particular to understand the not trivial $S$-shaped relaxation curves.

\subsection{Ground-state relaxation}
The hysteresis loops shown above, deduced from torque experiments in
the presence of relatively large longitudinal and transverse fields, give
evidence of temperature-independent tunneling below $\approx0.6$~K
(Fig.~\ref{fig4} and  Fig.~\ref{fig5}). Now we consider the relaxation
experiments performed in the same conditions of field and temperature, i.e.
around 4~T and at temperatures below 1.3~K. This will make a link with
previous experiments above 1.5~K, and fields up to 3~T. In both cases of
longitudinal and transverse field, the observed quantum relaxation can be very
fast and this allows to analyze in details the $M(t)$ curves even at
temperatures well below 1~K. 
\begin{figure}
\includegraphics[width=8.1cm]{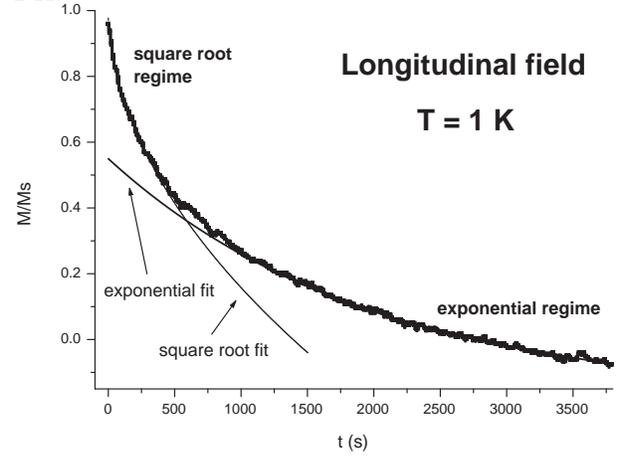}
\caption{Relaxation curve in a longitudinal field for the resonance $n=8$
($H\approx4$~T) and $T=1$~K. The square root regime is observed only at short
timescales, then the relaxation becomes exponential.} 
\label{fig9} 
\end{figure}
\begin{figure}
\includegraphics[width=8.1cm]{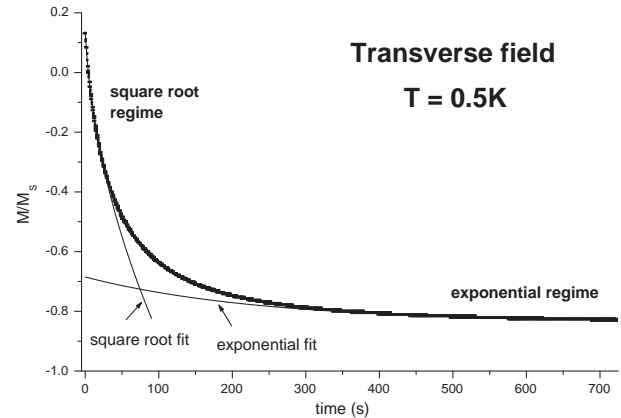}
\caption{Relaxation curve in a transverse field of 4~T for the resonance $n=0$
and $T=0.5$~K. The square root regime is observed only at short timescales,
then the relaxation becomes exponential.} 
\label{fig10} 
\end{figure}

\begin{figure}
\includegraphics[width=8.1cm]{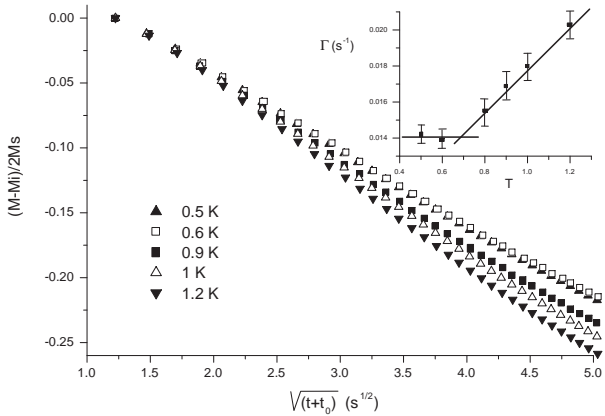}
\caption{Magnetic relaxation in Mn$_{12}$-ac for $n=0$ (transverse field), plotted vs. the
square root of time $t+t_0$ where $t_0=1.5$~s is imposed by our experimental
time resolution. The weak S-shaped character of the curve is due to small
shifts in the density of spin states with temperature (see section 4.2). In the
inset is plotted the deduced square root relaxation time for different
temperatures. One notes a plateau below $0.6-0.7$~K.}  
\label{fig11}  
\end{figure}

Two examples of relaxation curves measured in a longitudinal or a transverse
field are plotted in Fig.~\ref{fig9} and Fig.~\ref{fig10}. A square root
regime is clearly observed in both cases, but at short timescales only (see
also Fig.~\ref{fig11}). At longer timescales, the relaxation becomes
exponential in all cases. In these figures the square root and exponential
plots are done using the following expressions: 
\begin{equation} (M-M_i)/M_s =
\sqrt{\Gamma(t+t_0)} \end{equation} and 
\begin{equation}
(M-M_e) = A \exp{-(\Gamma^\prime t)} 
\end{equation}
Note that the data are much more noisy in longitudinal. This is because in
this case the torque is minimum (a very small transverse field component has
to be set to get a torque), while in the transverse case it is almost
maximum. Nevertheless, the square root and exponential character of the
relaxations is evident in both cases.

The square root relaxation time derived from short timescale experiments is
plotted in the inset of Fig.~\ref{fig10}. It is independent of temperature
below a crossover temperature $T_c$ of about $0.6-0.7$~K, showing
ground-state tunneling relaxation in the bulk phase of  Mn$_{12}$-ac for the
first time. The observed square root law, being for $n=0$, corresponds to
the switch of the total magnetization from $+M_s$ to $-M_s$  and it is
centered near $M=0$, i.e. for $M\ll M_s$, which was not predicted in
model\cite{ProkStamp}. It is interesting to note that the crossover
temperature $T_c(H)$ between ground-state and thermally assisted tunneling
does not seem to be related to the crossover temperature $T_b(H)$ between
square root and exponential relaxation. The reason is that the former depends
only on the temperature of the spin bath while the second depends also on the
equilibration of the spin-phonon system with the cryostat temperature.

Contrarily to the experiments performed at $T>2$~K and discussed above, the
square root laws observed below 1~K are not associated with the shift of the
total energy distribution of spins, but results from a hole burned by tunneling
in this energy distribution at zero energy, as predicted by Prokof'ev
and Stamp\cite{ProkStamp}. This hole must persist during all the square root
regime (about 30 seconds) and should vanish in the exponential regime which
characterizes an equilibrated spin system. The existence of such a finite hole
life-time below $T_c$ cannot be predicted in a zero Kelvin model. At finite
temperature the hole should vanish when the spin system becomes
equilibrated, i.e. when the tunnel window becomes as large as the width of
quasi-static internal field distribution. This is what is in principle
expected above $T_c$. The fact that this is observed even below $T_c$, at long
timescales, suggests that there are other mechanisms such as spin-phonon
transitions which reorganize the spin energy distribution at temperatures below
$T_c$.

Let us now evaluate the ground-state tunneling gap $\Delta$, from the
relaxation experiments in longitudinal and transverse fields of $\approx4$~T.

In the first case, the tunneling rate is given by $\Gamma =
\pi\Delta^2/(2\hbar E_D)$ where $E_D =(2S-n)g\mu_BH_D$. Knowing the
measured rate $\Gamma \approx10^{-3} s^{-1}$, $S=10$, $n=8$,
$g\mu_BH_D/k_B\approx0.05$~K, this expression gives $\Delta/k_B
\approx10^{-5}$~K. Interestingly the calculation of the splitting, by
diagonalization, using the anisotropy constants of  Mn$_{12}$-ac up to fourth
order, gives a value which is very close, $\Delta/k_B= 1.6\cdot10^{-5}$~K
(Table~\ref{table:1}).

In the second case, the tunneling rate is given by the same expression
where $E_D=2Sg\mu_BH_D$. Taking $g\mu_BH_D/k_B=0.05$~K, the measured rate
$\Gamma \approx10^{-2}$~s$^{-1}$ gives $\Delta/k_B \approx4\cdot10^{-5}$~K. The
diagonalization of the matrix gives $\Delta/k_B \approx10^{-5}$~K. Note
that when $\Gamma\tau_s\approx1$ the fast tunneling rate should be written
$[1- \exp{(-\pi\tau_s\Delta^2/ 2\hbar E_D)]}/\tau_s$. This expression
shows that $\tau_s$ plays effectively a role in the tunneling rate, but only
if $\Gamma\tau_s \approx1$. This is not the case in general, where slow
relaxation leads to the expression used above $\Gamma \approx
(\pi\Delta^2)/(2\hbar E_D)$. We conclude that ground-state relaxation is only
sensitive to changes in hyperfine fluctuations characteristic time. At
moderately high temperatures where $\tau_0 < \tau_s \approx (T_1,T_2)$, the
situation is different and $\tau_s$ might play a role in the relaxation as
shown in Eq.~\ref{GammaAs}. Finally, the slope $dn/dT_c$ in the inset of
Fig.~\ref{fig5} allows to evaluate the mean $\Delta$ for $5<n<11$. We find
$\Delta/k_B\approx1.5\cdot10^{-6}$~K.

Although much larger than in zero field, these gaps are still too small to be
perturbed by spin-phonon transitions. These transitions are relevant in the
thermally activated regime only. As discussed above, this restriction does not
apply to magnetic fluctuations of e.g. nuclear spins: a fluctuating field of
only $10^{-3}$~T would spread the Landau-Zener transition over 20~mK, i.e. 2000
times $\Delta$. This is why the spin bath, with fluctuations of amplitude $h_s$
and characteristic time $\tau_s$ ($c_s \approx h_s/\tau_s$), plays a dominant
role to induce tunneling. This is true unless $dH/dt$ is much
faster than hyperfine fluctuations. The condition for coherent adiabatic
Landau-Zener transition is $\Delta^2/\hbar g\mu_BS \gg c \gg c_s$. 

Finally, if there are no phonons with energy comparable to the tunneling gap
$\Delta \ll k_BT$, a large number of phonons must be available at the energy
scale of the hole width, and this explains why (i) non-equilibrated square root
behavior is observed even above $T_c$ at short timescales, (ii)
equilibrated exponential behavior is observed even below $T_c$ at long
timescales.

\section{Low spin systems (V$_{15}$ type)}

The process of spin reversal in this system has been studied by dynamical
magnetization experiments in the sub-Kelvin temperature range~\cite{ICprlV}.

In the molecular complex $K_6[V_{15}^{IV}As_6O_{42}(H_2O)]\cdot8H_2O$
(so-called V$_{15}$)~\cite{Muller} all intra-molecular exchange interactions
being antiferromagnetic, the total spin of the molecule is $S=1/2$; there is no
energy barrier and no tunneling. However the physics of spin rotation has some
similarities with the case of large spin molecules.

\begin{figure}
\includegraphics[width=8.1cm]{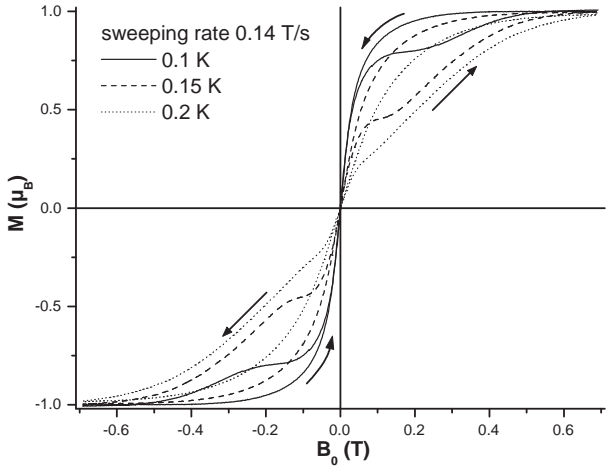}
\caption{Measured hysteresis loops for three temperatures and for a given
field sweeping rate 0.14~T/s. The plateau is more pronounced at low $T$.}
\label{fig12} 
\end{figure}

\begin{figure}
\includegraphics[width=8.1cm]{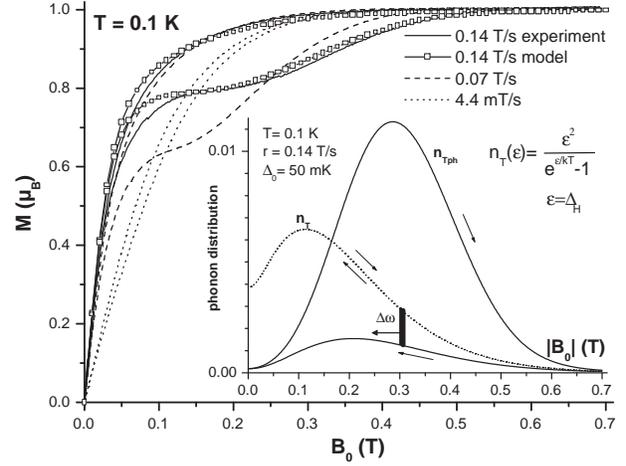}
\caption{Measured hysteresis loops for three field sweeping rates at
$T=0.1$~K. The observed plateau is more pronounced at high sweeping rate. The
square-symbol curve was calculated with the model discussed in text. We find a
good agreement between model and experiment. In the inset the calculated
number of phonons with $\hbar\omega \approx \Delta_0$ is plotted vs. the
sweeping field modulus (note the arrows) at equilibrium (dotted line) and
out-of-equilibrium (black line). The difference between the two curves (thick
segment $\Delta_0$) suggests the moving hole in the phonon distribution, while
their intersection gives the plateau intercept of the equilibrium magnetization
curve.}
\label{fig13}
\end{figure}

In Fig.~\ref{fig12} where three hysteresis loops are presented at three
different temperatures for a given sweeping rate, the plateau is higher and
more pronounced at low temperature. The same tendency is observed at a given
temperature and faster sweeping rate (Fig.~\ref{fig13}). When compared to its
equilibrium limit (dotted curve in Fig.~\ref{fig13}), each magnetization
curve shows a striking feature: the plateau intersects the equilibrium curve
and the magnetization becomes smaller than at equilibrium. Equilibrium is then
reached in higher fields near saturation.

In order to interpret this magnetic behavior of the V$_{15}$ molecules, we
will analyse how the level occupation numbers vary in this two level system
when sweeping an external field. In the absence of dissipation, a 2-level
model is well described by the bare Landau-Zener model, in the adiabatic or
non-adiabatic case (low or high sweeping rates).  The probability for the
$|1/2,-1/2\rangle\leftrightarrow|1/2,1/2\rangle$ transition is
$P=1-\exp(-\pi\Delta_0^2/4\hbar\mu_Bc)$. In such a Landau-Zener  transition,
the plateaus of Fig.~\ref{fig12},\ref{fig13} should decrease if the sweeping
rate increases, which is contrary to the experiments. Taking the typical value
$c=0.1$~T/s and the zero-field splitting $\Delta_0/k_B\approx 0.05$~K
~\cite{ICprlV}, one gets a ground state switching probability very close to
unity: in the absence of dissipation the spin 1/2 must adiabatically follow
the field changes. But, dissipation due to spin-phonon coupling make the
transition dissipative (or non-adiabatic, from the thermodynamical point of
view). The mark of the $V_{15}$ system is that this coupling is acting also
near zero applied field because $\hbar\omega\approx\Delta_0$ is of the order
of the bath temperature, which is not the case of large spin molecules where
$\Delta_0 \ll k_BT$.

In a direct process, the spins at the temperature $T_S$ should relax to
the phonon temperature within a timescale $\tau_1$, the phonons being at the
bath temperature. However, even with a silver sample holder, it is not
possible to maintain the phonon temperature equal to the temperature of the
bath. This is because in V$_{15}$ below 0.5 K, the heat capacity of the
phonons $C_{ph}$ is very much smaller than that of the spins $C_S$, so that
the energy exchanged between spins and phonons will very rapidly adjust the
phonon temperature $T_{ph}$ to the spin one $T_S$. Furthermore, the energy
is transferred from the spins only to those phonon modes with
$\hbar\omega=\Delta_H$ (within the resonance line width), where $\Delta_H$ is
the two level field-dependent separation. The number of such lattice modes
being much smaller than the number of spins, energy transfer between the
phonons and the sample holder must be very difficult, a phenomenon known as
the phonon bottleneck ~\cite{VVleckStev}. If this phonon density of states does not
equilibrate fast enough, the hole must persist and move with the sweeping
field, leading to a phonon bottleneck. This means that the spin system will
not be able to get the equilibrium, as soon as an external field is swept with
a sufficiently large rate. Let us note that in zero field the system is
out-of-equilibrium even if magnetization passes through the origin of
coordinates. At larger fields, in the plateau region, the level population is
almost constant at timescales shorter than $\tau_H$, even after the plateau
crosses the equilibrium curve. Equilibrium is reached when $\tau_H$ becomes
small enough.

In~\cite{ICprlV} we give a theoretical model inspired from the well-known work
of Abragam \& Bleaney ~\cite{AB}. Starting from the balance equations of
the coupled spin-phonon two level system, we derive the relaxation law of the
magnetic moment: $-t/\tau_H=x(t)-x(0)+\ln((x(t)-1)/(x(0)-1))$, where $x(t)$ is
the deviation of the magnetization from its equilibrium. $\tau_H$ is the
relaxation time and is a function of field and temperature (see
~\cite{ICprlV,AB}). We then calculate the hysteresis loops with very
realistic values of the parameters and we find a good agreement between
measurements and the calculus of hysteresis loops and relaxation times.

The V$_{15}$ molecular complex constitutes an example of
dissipative two-level system of mesoscopic size. The total spin 1/2 being
formed of a large number of interacting spins, its splitting results from the
structure itself of the molecule (intra-molecular hyperfine and
Dzyaloshinsky-Moriya couplings) and it is rather large (a fraction of Kelvin).
In V$_{15}$ and in other low-spin systems, splittings must be much larger than
in large-spin molecules where the presence of energy barriers lowers them by
orders of magnitude (e.g. $10^{-11}$~K in Mn$_{12}$, see above). This is the
reason why spin-phonon transitions are important in low-spin molecules and not
relevant in high-spin ones, unless a large transverse field is applied (it
increases the tunnel splitting and probability) in which case we would also
expect similar phenomena. 

\section{Conclusion}
The evidence of resonant tunneling of the magnetization in Mn$_{12}$-ac allows
to make a link between magnetism and mesoscopic physics. In particular it
shows that the transition between the low temperature quantum behavior and the
high temperature classical one, takes place through an intermediate regime in
which tunneling is thermally assisted. These studies also shed light on the
effects of spin couplings with their environment. The magnetic relaxation law
$M(t)$ and resonance line-shape $dM/dH$ are connected to each other. They both
give different results depending on the strength of the coupling with the bath
: square root relaxation and any type of line-shape when the spin (and phonon)
system is not at equilibrium, and exponential relaxation with Lorentzian
line-shape when it is equilibrated. However, this is not an absolute
rule and we found, and explained, situations where a square root
relaxation can be also observed when the spin system is at equilibrium.	These
results obtained previously above 1.5~K are confirmed by high fields sub-kelvin
experiments. In particular the crossover between square-root and exponential
relaxation law at $T_b$, and the crossover between ground-state and thermally
activated tunneling at $T_c$, are clearly observed in the bulk phase of
Mn$_{12}$-ac.  Note that these two crossovers do not coincide. This
observation is not surprising because the former is related to different
couplings with the bath (spins, phonons), while the latter is essentially a
single molecule effect. Finally, we have evaluated the tunneling gaps
of the bulk phase of Mn$_{12}$-ac for different experimental cases. A very
good agreement was always obtained with the gaps obtained by exact
diagonalization.

The role of spin-phonon transitions is discussed in details
in the low spin molecule V$_{15}$, where the absence of barrier leads to large
gaps. This study, in which the reversal of a ``macroscopic" spin $1/2$ is
analyzed, shows particular hysteresis loops reminiscent but definitely
different from those of high spin molecules. The hysteresis loop of low spin
molecules comes from spin-phonon transitions. It is concluded that in the
presence of a large magnetic field, spin-phonon transitions should also be 
important in large spin molecules and lead to hysteresis loops similar to
these observed in V$_{15}$.

\section*{Acknowledgements}
We thank T. Goto, S. Maegawa, S. Miyashita, N. Prokof'ev, P. Stamp, I.
Tupitsyn and A. Zvezdin,, for fruitful on-going collaboration and
discussions.


\begin{thebibliography}{99}

\bibitem{MUBB} M. Uehara and B. Barbara: J. Phys. {\bf 47} (1986) 235 and
refs. therein.    

\bibitem{LGBB} L. Gunter, B. Barbara (Eds.): Quantum Tunneling of
Magnetization, NATO ASI series E: Applied Sciences {\bf 301}, Kluwer,
Dordrecht (1995).     

\bibitem{Egami} T. Egami: Phys. Stat. Sol. B {\bf 57} (1973) 211; ibid. A {\bf
19} (1973) 747; {\bf 20} (1973) 57.   

\bibitem{BB} B.  Barbara: Proc. 2$^{nd}$ Int. Symp. on Anisotropy and
Coercivity {\bf 137} (1978); J. de Phys. {\bf 34} (1977) 1039; Sol.
State Com. {\bf 10} (1972) 1149 .    

\bibitem{Paulsen} C. Paulsen, J. G. Park, B. Barbara, R. Sessoli, A. Caneschi:
J. Magn. Magn. Mat. {\bf 140-144} (1995) 1891.    

\bibitem{Paulsen1} C. Paulsen, J. G. Park, B. Barbara, R. Sessoli, A. Caneschi:
J. Magn. Magn. Mat. {\bf 140-144} (1995) 379. 

\bibitem{BBjmmm95} B. Barbara et al.: J. Magn. Magn. Mater. {\bf 140-144} (1995) 1825.     

\bibitem{Paulsen2} C. Paulsen and J. G. Park: in \cite{LGBB} p. 189. 
 
\bibitem{Novak} M. A. Novak and R. Sessoli: in \cite{LGBB} p. 171. 
 
\bibitem{Fried} J. R. Friedman, M. P. Sarachik, J. Tejada, R. Ziolo: Phys.
Rev. Lett. {\bf 76} (1996) 3830. J. M. Hernandez, X. X. Zhang, F. Luis,
J. Bartolome, J. Tejada, R. Ziolo: Europhys. Lett. {\bf 35} (1996) 301.   

\bibitem{Luc} L. Thomas, F. Lionti, R. Ballou, R. Sessoli, D. Gatteschi and 
B. Barbara: Nature {\bf 383} (1996) 145. 

\bibitem{Peren} J. A. A. J. Perenboom, J. S. Brooks, S. Hill, T. Hathaway and 
N. S. Dalal: Phys. Rev. B {\bf 58} (1998) 333.   

\bibitem{jmmm200} B. Barbara, L. Thomas, F. Lionti, I. Chiorescu, A. Sulpice:
J. Magn. Magn. Mat. {\bf 200} (1999) 167.   

\bibitem{IC} I. Chiorescu, R. Giraud, A. Caneschi, A. G. M. Jansen,
B. Barbara: to be published.   

\bibitem{BBjmmm98} B. Barbara, L. Thomas, F. Lionti, A. Sulpice, A. Caneschi:
J. Magn. Magn. Mat. {\bf 177-181} (1998) 1324.   

\bibitem{LucPRL} L. Thomas, A. Caneschi, B. Barbara: Phys. Rev. Lett. {\bf 83} (1999) 2398.   

\bibitem{FredJAP} F. Lionti, L. Thomas, R. Ballou, B. Barbara, A.
Sulpice, R. Sessoli, D. Gatteschi: J. Appl. Phys. {\bf 81} (1997) 4608. 
 
\bibitem{BarraPRB} A. L. Barra, D. Gatteschi, R. Sessoli: Phys. Rev. B {\bf 56} (1997) 8192 .   

\bibitem{Hemmen} J. L. van Hemmen and A. S\"ut\H o: in \cite{LGBB} pg. 19.

\bibitem{Miyashita} S. Miyashita: J. Phys. Soc. Jap. {\bf 64} (1995) 3207;
J. Phys. Soc. Jap. {\bf 65} (1996) 2734.   

\bibitem{LZ}  L. D. Landau: Phys. Z. Sowjetunion {\bf 2} (1932) 46.
C. Zener: Proc. R. Soc. London A {\bf 137} (1932) 696.   

\bibitem{LGunther} L. Gunther: EuroPhys. Lett. {\bf 39} (1997). 
 
\bibitem{Dobro} V. V. Dobrovitski, A. K. Zvezdin: Europhys. Lett. {\bf 33}
(1997) 377.      
\bibitem{Garanin} D. A. Garanin and E. M. Chudnovsky: Phys. Rev. B {\bf 56} (1997) 11102.    

\bibitem{Igor} I. Tupitsyn, B. Barbara: to be published.	   

\bibitem{Leggett} A. J. Leggett et al.: Rev. Mod. Phys. {\bf 59} (1987) 1;
Lectures in Phys., Les Houches (1986).  A. J. Leggett: in \cite{LGBB} p. 1;
A. O. Caldeira and A. J. Leggett: Phys. Rev. Lett. {\bf 46} (1981) 211. P.
C. E. Stamp: {\bf 61} (1988) 2905; N. Prokof'ev and P. C. E. Stamp: J. of
Low Temp. Phys. {\bf 104} (1996) 143; J. of Phys. Cond. Matt. {\bf 5} (1993) L663.

\bibitem{Stamp1} P. C. E. Stamp, E. Chudnovsky and B. Barbara: Int. J. of Mod.
Phys. B {\bf 9} (1992) 1355.    

\bibitem{StampPRL88} P. C. E. Stamp: Phys. Rev. Lett. {\bf 61} (1988) 2905.    

\bibitem{Berry} P. M. V. Berry: Proc. R. Soc. London A {\bf 392} (1984) 45.

\bibitem{ProkStamp} N. V. Prokof'ev and P. C. E. Stamp: Phys. Rev. Lett. {\bf
80} (1998) 5794 and J. of Low Temp. Phys. {\bf 113} (1998) 1147.    

\bibitem{Villain} J. Villain, F. Hartmann-Boutron, R. Sessoli, A. Rettori:
Europhys. Lett. {\bf 27} (1994) 159; P. Politi, A. Rettori,
F. Hartmann-Boutron, J. Villain: Phys. Rev. Lett. {\bf 75} (1995) 537;
F. Hartmann-Boutron, P. Politi, J. Villain: Int. J. Mod. Phys. B {\bf 10}
(1996) 2577; M. Leuenberger and D. Loss: Europhys. Lett. {\bf 46} (1999) 692.
   

\bibitem{Goto} T. Goto, T. Kubo, T. Koshiba, Y. Fujii, A. Oyamado, J. Arai,
K. Takeda, K. Awaga: preprint sep.1999, to appear in Physica B (2000).    

\bibitem{Borsa} A. Lascialfari, Z. H. Jang, F. Borsa, P. Carretta and
D. Gatteschi: Phys. Rev. Lett. {\bf 82} (1998) 3773.    

\bibitem{Ohm} T. Ohm, C. Sangregorio and C. Paulsen: J. of Low Temp. Phys.
{\bf 113} (1998) 1141.

\bibitem{LucLowT} L. Thomas and B. Barbara: J. of Low Temp. Phys. {\bf 113}
(1998) 1055.

\bibitem{ICprlV} I. Chiorescu, W. Wernsdorfer, A. M\"uller, H. B\"ogge and B.
Barbara: Phys. Rev. Lett. {\bf 84} no.15 (2000). 
  
\bibitem{Muller}  A. M\"uller, J. D\"oring: Angew. Chem. Int. Ed.
Engl. {\bf 27} (1991) 1721. D. Gatteschi, L. Pardi, A. L. Barra, A.
M\"uller, J. D\"oring: Nature {\bf 354} (1991) 465. 

\bibitem{VVleckStev} J. H. Van Vleck: Phys. Rev. {\bf 59} (1941) 724. K. W. H.
Stevens: Rep. Prog. Phys. {\bf 30} (1967) 189.   

\bibitem{AB} A. Abragam, B. Bleaney: Electronic Paramagnetic
Resonance of Transitions Ions, 	Clarendon Press - Oxford, chap. 10, (1970).   

\bibitem{Kent} A. D. Kent, Y. Zhong, L. Bokacheva, D. Ruiz, D. N.
Hendrickson, M. P. Sarachik: EuroPhys. Lett. {\bf 49} (4) (2000) 521. 

\end{thebibliography}
\end{document}